\documentclass[twocolumn,prl,superscriptaddress,a4paper]{revtex4}

\usepackage{amsmath,amssymb,graphicx}

\begin{document}

\setlength{\tabcolsep}{1.5mm}
\setcounter{totalnumber}{4}
\setcounter{topnumber}{4}

\setlength{\voffset}{-0cm}
\setlength{\hoffset}{-0.cm}
\addtolength{\textheight}{1.1cm}

\title{Self-templated nucleation in peptide and protein aggregation}

\author{Stefan Auer}
\affiliation{Centre for Self Organising Molecular Systems, University of Leeds,
Leeds LS2 9JT, United Kingdom}

\author{Christopher M.~Dobson}
\affiliation{Department of Chemistry, University of Cambridge, Lensfield Road,
Cambridge CB2 1EW, United Kingdom}

\author{Michele Vendruscolo}
\affiliation{Department of Chemistry, University of Cambridge, Lensfield Road,
Cambridge CB2 1EW, United Kingdom}

\author{Amos Maritan} 
\affiliation{Dipartimento di Fisica, Universit\'a di Padova, 
INFN and CNISM, Via Marzolo 8, 35131 Padova,Italy}


\date{\today}

\begin{abstract}
  Peptides and proteins exhibit a common tendency to assemble into
  highly ordered fibrillar aggregates, whose formation proceeds in a
  nucleation-dependent manner that is often preceded by the formation
  of disordered oligomeric assemblies. This process has received much
  attention because disordered oligomeric aggregates have been
  associated with neurodegenerative disorders such as Alzheimer's and
  Parkinson's diseases. Here we describe a self-templated nucleation
  mechanism that determines the transition between the initial
  condensation of polypeptide chains into disordered assemblies and
  their reordering into fibrillar structures. The results that we
  present show that at the molecular level this transition is due to
  the ability of polypeptide chains to reorder within oligomers into
  fibrillar assemblies whose surfaces act as templates that stabilise
  the disordered assemblies.
\end{abstract}

\maketitle

There are two fundamental questions that one can ask about the general
phenomenon of formation of ordered structures by atoms and molecules.
The first concerns the type of assembly that the given particles can
form, and the second the kinetic paths that the particles follow in
order to reach a stable structure.  The answer to the first question,
at least in the cases when the particles are rigid, is that in nature
there exists only a small number of possible crystal structures,
corresponding to the $230$ crystallographic space groups
\cite{cornwell84}.  The extraordinary power of this result, which is
based on symmetry and geometry arguments, is that the answer is
independent of the specific particles and of their mutual
interactions.  The specific particle properties merely determine which
type of crystal structure is the most stable.  The answer to the
second question is generally more difficult, and involves a nucleation
and growth mechanism \cite{kelton91}. In this mechanism the atoms
first need to come together to form a critical nucleus (nucleation
phase), before they can grow (elongation phase).  The probability
$P_c$ that a spontaneous fluctuation will result in the formation of a
critical nucleus depends exponentially on the free energy $\Delta
F_c$ required to form such a nucleus: $P_c=\exp(-\Delta F_c/kT)$,
where $T$ is the absolute temperature and $k$ is the Boltzmann
constant. In atomic systems the activation barriers are usually very
high, and the probability of observing nuclei is very small; even when
they form, their lifetime is fleetingly short, so that up to now there
is no clear experimental observation of critical nuclei in atomic
systems, and computer simulations have become a major tool to
investigate this phenomenon \cite{sear07}.

An additional problem arises when the particles forming the ordered
structures are not rigid, but flexible, as is the case of peptides and
proteins. Individual molecules of this type have often an intrinsic
tendency to fold into ordered structural patterns, which may either
favour or hinder their intermolecular assembly process.  This problem
constitutes an entirely new chapter in the study of ordering that has
very great significance for biology and biotechnology.  Indeed
biomolecules such as DNA and proteins have recurrent structural motifs
such as $\alpha$-helices and $\beta$-sheets \cite{alberts02}, and a
wide range of different proteins can assemble into highly ordered
fibrillar aggregates \cite{chiti06}.  Although the amino acid sequences
of these proteins are often unrelated, the structures of amyloid
fibrils show a common characteristic cross-$\beta$ structure in which
the main axis of individual molecules runs orthogonal to the direction
of the filaments. It has thus been suggested that the inherent ability
to form fibrillar assemblies is a feature common to polypeptide
chains \cite{dobson03a}.

\begin{figure}
\includegraphics[width=89mm]{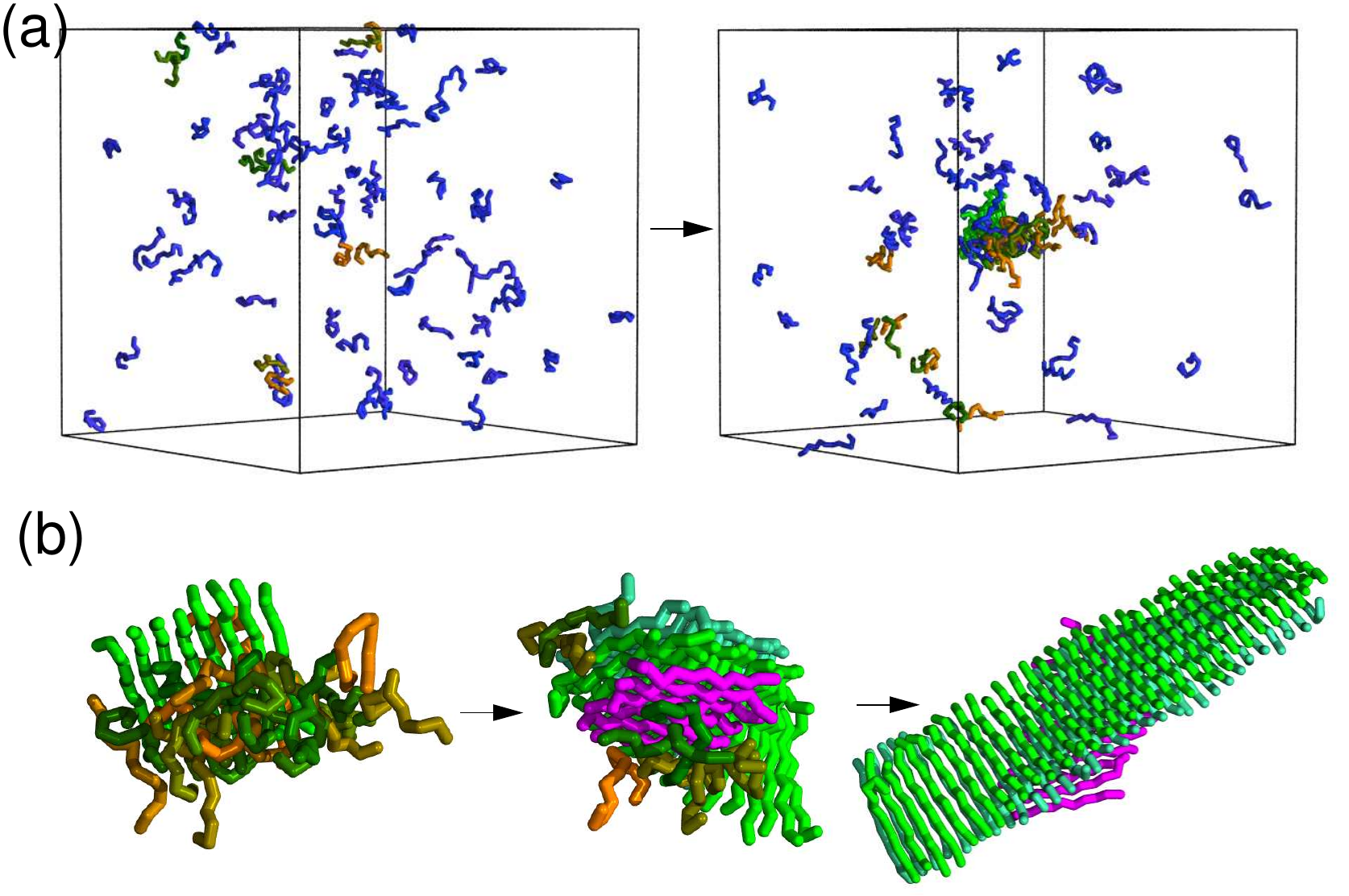}
\caption{
\label{fig:snapshot}
Condensation-reordering mechanism at $c=12.5mM$ and $T^*=0.66$ above
the folding temperature. Peptides that do not form interchain hydrogen
bonds are shown in blue, those forming interchain hydrogen bonds are
assigned a random colour. Peptides within the same $\beta$-sheet are
assigned the same colour. (a) Initially, at $t\leq 1000$, the peptides
are in a monomeric state (left panel). The progress variable $t$ is
the number of Monte Carlo moves performed in the simulation, and one
unit of $t$ is a block of $10^5$ Monte Carlo moves. As the simulation
progresses, a hydrophobic collapse causes the formation of a small
disordered oligomer ($t = 15000$, right panel). (b) Enlarged view of a
disordered oligomer which subsequently orders into a protofilament
structure: $t = 15000$ (left), $t = 19000$ (middle), $t = 30000$
(right).}
\end{figure}

In order to explain why this process takes place despite the
remarkable resistance of native states of proteins to aggregation, a
``nucleated conformational conversion" mechanism has been proposed in
which the formation of highly dynamic oligomeric assemblies
facilitates the further conversion of polypeptide chains into ordered
fibrillar structures \cite{serio00}.  Evidence in favour of this
mechanism has been provided through experimental
\cite{nelson05,tanaka06,lomakin96} and theoretical studies
\cite{urbanc04,nguyen04,ma06,hills07,pellarin06,cheon07}.  In this
work we exploit the ability of computer simulations to provide a
description of molecular process at very high resolution - an ability
that has proven invaluable in defining nucleation processes in atomic
and colloidal systems \cite{auer01,sear07}. In the case of polypeptide
chains we have recently shown that it is possible to follow the
aggregation process on a time long enough to include both the initial
condensation into disordered oligomeric assemblies and their
subsequent reorganisation into fibrillar structures \cite{auer07}, and
that the nucleated conformational conversion mechanism can also be
described as a ``condensation-reordering" mechanism
(Fig.~\ref{fig:snapshot}). Here we investigate a self-templated
nucleation mechanism, which determines the coupling between the
assembly of polypeptide chains in disordered oligomers and the
transformation into highly ordered cross-$\beta$ structures.  A
template model by which misfolded prions induce the conversion of
nearby native prions to the misfolded state has already been proposed
\cite{prusiner98}, and described for SH3 \cite{ding02} and cc$\beta$
\cite{ding05}.  In particular, in the latter case it was shown that
$\beta$-sheets with exposed hydrophobic surfaces and unsaturated
hydrogen bonds accelerate the conversion from native $\alpha$-helices
to $\beta$-sheets.

The computational strategy that we follow is based on the attempt to
reproduce simultaneously two common aspects of proteins, their ability
to form secondary structural motifs \cite{pauling51}, and their
propensity to form ordered fibrillar aggregates \cite{chiti06}.
Recently it has be shown that motifs such as $\alpha$ helices, $\beta$
sheets and cross-$\beta$ structures are natural forms of a marginally
compact phase of matter characteristic of flexible
polymers \cite{banavar04,banavar07}. The description of proteins in terms of
flexible tubes captures in a simple way the main symmetry of chain
molecules.  In this approach a polypeptide chain is represented by a
$C_{\alpha}$ chain of finite thickness, which approximately envelops
the backbone atoms. The hydrophobic effects due to the water are
considered by a pairwise additive interaction between different
$C_{\alpha}$ atoms, with an energy $e_{HP}$, when they are close. The
sequence independent definition for hydrogen bonding is obtained by an
analysis of the geometric properties of hydrogen bond forming
$C_{\alpha}$ atoms from the Protein Data Bank and assigned an energy
$e_{HB}$. Steric constraints due to side chains are imposed by local
bending stiffness with energy $e_{S}$ (for a detailed description of
the model see \cite{hoang04,auer07}). Based on the
hypothesis that the formation of amyloid fibrils is a common feature
of all polypeptide chains, which depends mainly on the generic
properties of their backbone \cite{dobson99a,knowles07}, we
investigated the behaviour of a representative model system consisting
of 80 weakly hydrophobic 12-residue homopolymers. Systems of
homopolymers \cite{aggeli01} have been shown experimentally to form
amyloid assemblies. In all our simulations the energy of hydrogen
bonds was set to $e_{\rm HB}=-3kT_o$, a value close to experiments
($1.5$kCal/mol at room temperature \cite{fersht85}). Here $kT_o$ is a
reference thermal energy and the reduced temperature is $T^*=T/T_o$.
The hydrophobic and stiffness energy are set to $e_{HP}=-0.15kT_o$ and
$e_S=0.9kT_o$ respectively. The ratio $e_{\rm HB}/e_{\rm HP}=20$ is
such that these interactions provide similar contributions to the
potential energy of the oligomer. With this choice of parameter the
peptides form an $\alpha$ helical native structure below the folding
temperature $T^*_f\sim 0.6$, and a random coil above.

\begin{figure*}
\includegraphics[width=164mm]{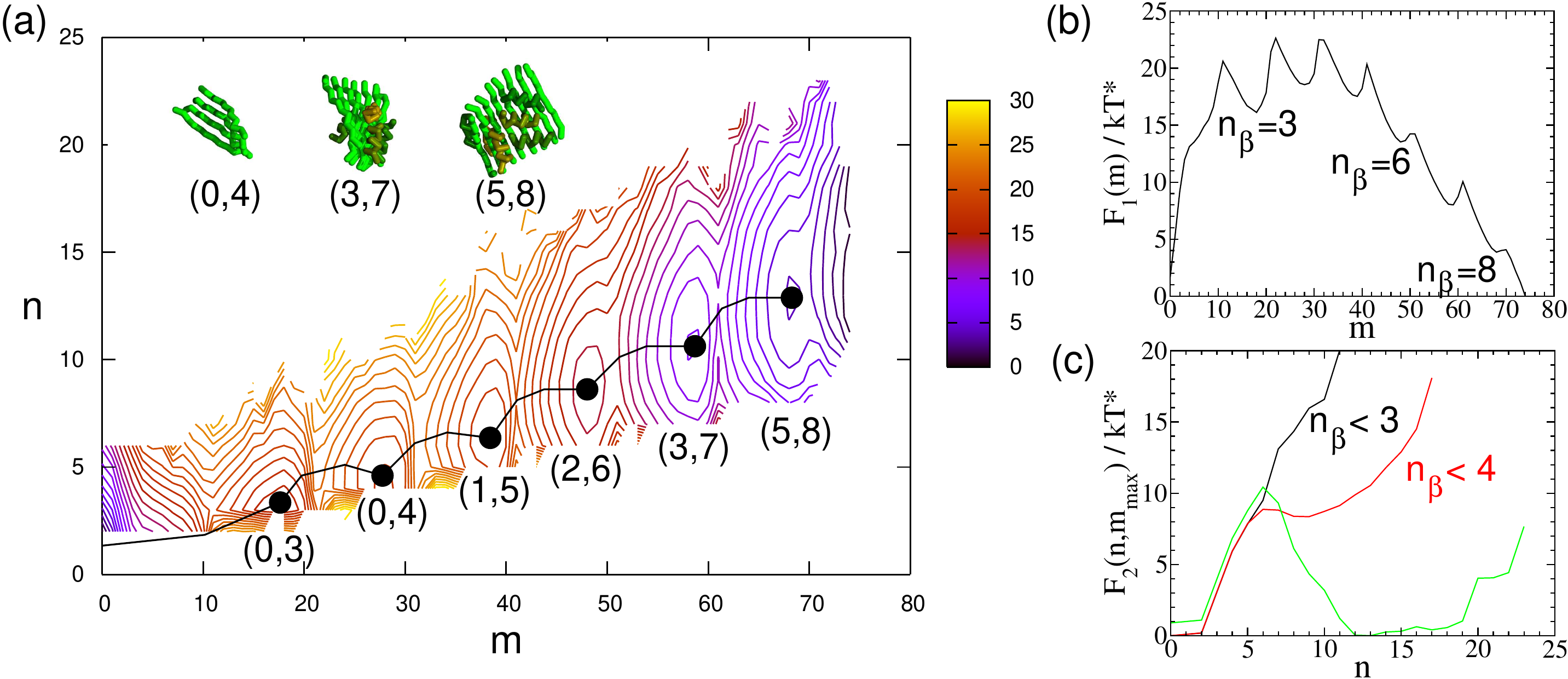}
\caption{
\label{fig:contour}
(a) Contour plot of the nucleation barrier $F(n,m)$ calculated at
concentration $c=1.2mM$ and reduced temperature $T^*=0.51$, which is
below the folding temperature.
The black circles indicate the minima on the
free energy surface, and the black line indicates a possible path that
connects them. The labels $(n_{\alpha},n_{\beta})$ of the minima
describe the structure of the oligomer, where $n_{\alpha}$ and
$n_{\beta}$ are, respectively, the number of peptides in a
$\alpha$-helical and $\beta$-strand structure, and $n=n_{\alpha} +
n_{\beta}$. In the inset we show snapshots of the oligomers associated
with the minima. (b) Nucleation barrier for $\beta$-sheet formation,
$F_1(m)$, as a function of the number of interchain hydrogen bonds $m$
that are formed within the oligomer. (c) Nucleation barrier
$F_2(n,m_{max})$ for the formation of an oligomer of size $n$ that can
at most form $m_{max}$ interchain hydrogen bonds: $m_{max}=10$
($n_{\beta}<3$) (black line), $m_{max}=20$ ($n_{\beta}<4$) (red line),
all $m$ values included (green line).}
\end{figure*}

In order to illustrate the condensation-reordering transition
(Fig.~\ref{fig:snapshot}) we set the peptide concentration to
$c=12.5mM$ and the reduce temperature to $T^*=0.66$, to keep the lag
time prior to aggregation very short.  Lowering the concentration,
while keeping the temperature constant, results in a dramatic increase
of the lag time.  At $c=1mM$ the peptides remain monomeric on the
timescale that we have been able to follow, although the aggregated
state is likely to be much more stable than the monomeric state.  In
this concentration regime ($c=1mM$ to $c=12.5mM$) the monomeric state
is metastable with respect to the aggregated state, and the
aggregation of polypeptide chains follows a nucleation
mechanism \cite{jarrett93}. Under such conditions we calculated the
nucleation barriers associated with the condensation-reordering
transition.

\begin{figure}
\includegraphics[width=82mm]{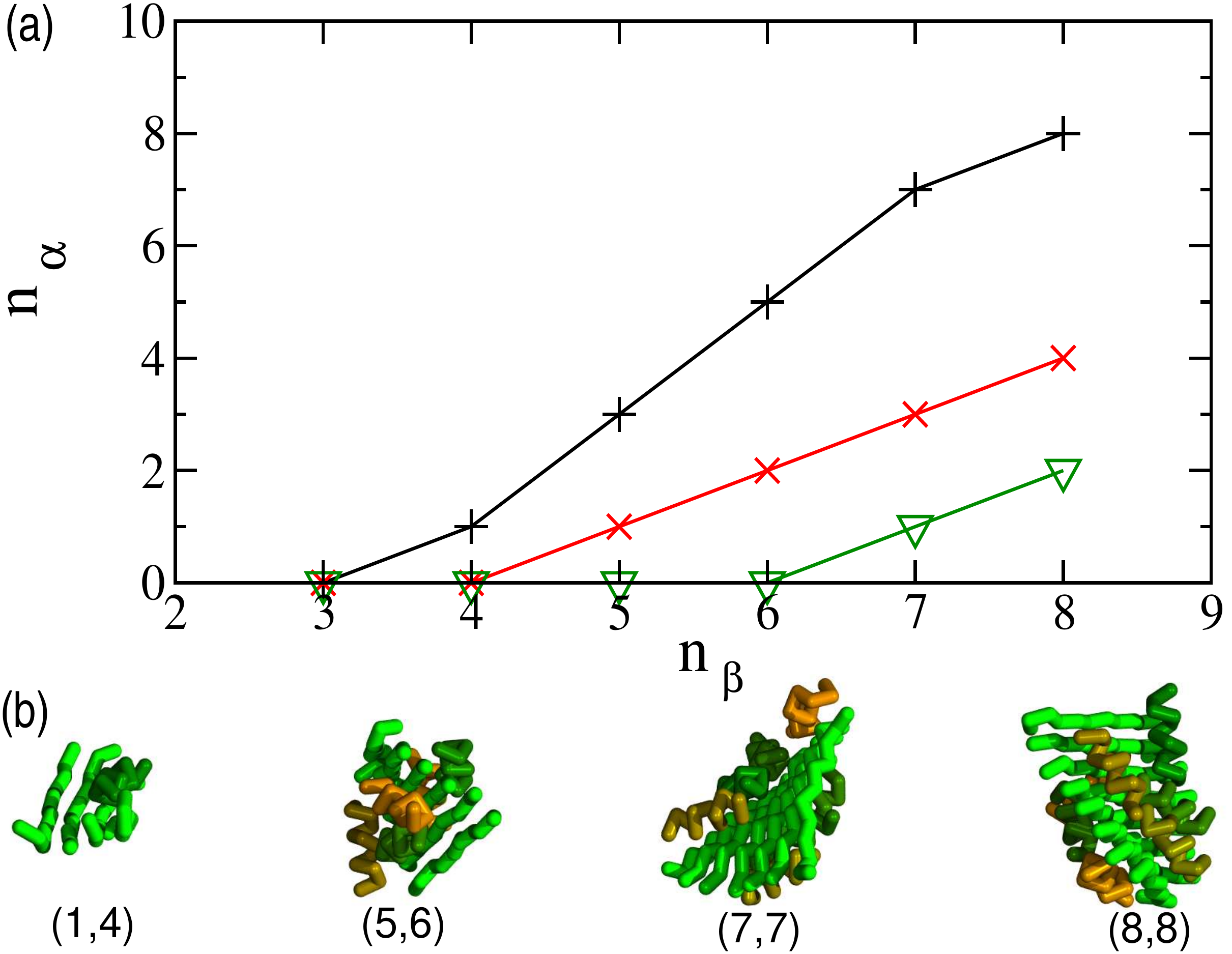}
\caption{ Correlations for the successive minima in the free energy
  landscape (such as shown in Fig.  \ref{fig:contour}a) between the
  number $n_\alpha$ of peptides in a $\alpha$-helical conformation and
  the number $n_\beta$ of peptides in an $\beta$-sheet conformation.
  (a) Correlations obtained at $T^*=0.51$ and $c=0.64mM$ (green line),
  $c=1.2mM$ (red line) and $c=2.9mM$ (black line). (b) Typical
  configurations of the oligomers associated with the minima obtained
  at $T^*=0.51$ and $c=2.9mM$ (black line in
  (a)).\label{fig:correlation}}
\end{figure}

A prerequisite for such a calculation is the ability to describe
quantitatively the formation of small oligomeric assemblies. By using
a standard cluster criterion, i.e. any two peptides whose centre of
mass distance is less than $5\AA$ belong to the same oligomer, it is
possible to define the oligomer size $n$, that corresponds to the
number of peptides within the oligomer.  At the same time it is
possible to measure the $\beta$-sheet content of the oligomer. As the
formation of $\beta$-sheets is driven by interchain hydrogen bonding,
we use $m$, the number of interchain hydrogen bonds formed within an
oligomer, as a structural observable. In order to calculate the joint
equilibrium probability $P(n,m)$ for the formation of an oligomeric
assembly consisting of $n$ peptides with $m$ interchain hydrogen bonds
we performed biased Monte Carlo simulations \cite{auer01,auer07} in
the canonical ensemble using crankshaft, pivot, reptation, rotation,
and translation moves.  As a biasing potential we included an
additional parabolic energy term, $W = \alpha (m-m_0)^2$, in the
energy function, where $\alpha$ and $m_0$ are parameters, that can be
used to control the range of $m$ values sampled in the simulation. The
calculation of $P(n,m)$ was split into $28$ independent simulations
for different $m_{0}$ values, and finally combined into one by a
multi-histogram technique. The corresponding nucleation free-energy
landscape, apart from an additive constant, is given by
$F(n,m) = - kT^*\ln[P(n,m)].$
A representative calculation of such a free energy landscape obtained
at $c=1.2mM$ and $T^*=0.51$ (i.e. below $T^*_f$) is shown in
Fig.~\ref{fig:contour}a. The succession of local minima in this
landscape reveals the central result of this work - the presence of a
coupling between the oligomer size $n$ and the number of hydrogen
bonds $m$ in the states of minimal free energy for the oligomers.  To
clarify the nature of this coupling we analysed the structure of the
oligomers formed in the local minima of the free energy landscape.  As
an example, we show in the inset of Fig.~\ref{fig:contour}a that the
oligomer corresponding to the minimum $(n=13,m=68)$ consists of
$n_{\beta}=8$ peptides in a $\beta$ sheet conformation and
$n_{\alpha}=5$ in a helical conformation. Here we define two peptides
to be part of a $\beta$-sheet if they form more than four interchain
hydrogen bonds with each other. Since the peptides within the oligomer
are either in a $\alpha$-helical or $\beta$-strand conformation,
$n=n_{\alpha}+n_{\beta}$, we labelled each minimum by the pair
$(n_{\alpha},n_{\beta})$. Typical configurations of other oligomers
are also shown in the inset of Fig.~\ref{fig:contour}a. We further
analyse this coupling in Fig.~\ref{fig:correlation}, where we plot,
for the states corresponding to the minima of the free energy, the
number $n_\alpha$ of peptides in an $\alpha$-helical conformation
within an oligomer as a function of the number $n_\beta$ of peptides
in a $\beta$-sheet conformation. The essentially linear relationship
indicates that the probability of oligomeric assemblies
increases with the size of the ordered $\beta$-sheet structures.

To reveal the molecular basis of this coupling, we have analyzed the
nucleation barrier for $\beta$-sheet formation independent of the
number of peptides in a $\alpha$-helical conformation. The average
over $n$ is achieved by the marginalisation of $P(n,m)$ with respect
to $n$, and its corresponding free-energy profile is
$F_1(m)= -kT^*\ln[\sum_n P(n,m)].$
The calculations indicate that $F_1(m)$ is comprised of a series of
component barriers, each separated by $\Delta m=10$
(Fig.~\ref{fig:contour}b). Since each pair of peptides considered
here in a $\beta$-sheet conformation can form at most ten interchain
hydrogen bonds with each other, each maximum of $F_1(m)$ corresponds
to the addition of a new peptide to the existing $\beta$-sheet
structure within the oligomer. After the first few interchain hydrogen
bonds are formed, the free energy decreases until an optimal number of
hydrogen bonds is formed, corresponding to a local minimum. The
elongation barrier for $\beta$-sheet nucleation corresponds to the
free energy needed to transform a peptide from its $\alpha$-helical
conformation to the extended structure it has in its $\beta$-sheet
conformation. The elongation barrier is a quantitative measure for the
aggregation propensities of proteins, and can be measured
experimentally \cite{knowles07}. In our calculation a dimer formed by
two $\beta$-strands is always unstable and disassembles soon after it
forms; the critical size of the $\beta$-sheet is a tetramer, at least
in the range of concentrations and temperatures that we considered.

Next we investigated how the nucleation barrier for the formation of
an oligomer depends on its internal structure, i.e. if a $\beta$-sheet
is present or not. The structural average over $m$ is obtained by
resorting to the marginal distribution function $P(n,m\leq
m_{max})\equiv \sum_{m=0}^{m_{max}} P(n,m)$ and the corresponding
free-energy profile is given by
$F_2(n,m_{max})= -kT^* \ln P(n,m\leq m_{max}).$
Here $F_2(n,m_{max})$ is the nucleation barrier for forming an
oligomer of $n$ peptides, with at most $m_{max}$ interchain hydrogen
bonds. This upper limit to the number of interchain hydrogen bonds is
introduced in the projection operation with the goal of unveiling the
role of the $\beta$-sheet structure formed within the oligomer in the
dynamical evolution of the aggregation process. If we do not allow the
formation of $\beta$-sheets consisting of more than two peptides, by
imposing $m_{max} = 10$, the free energy for the formation of an
oligomer increases monotonically (Fig.~\ref{fig:contour}c, black
line). If we do allow the formation of $\beta$-sheets consisting of up
to three peptides, by relaxing the constrain to $m_{max} = 20$, we
observe that after an initial increase, for $n<5$, a local minimum is
present at $n_{min}=10$ (Fig.~\ref{fig:contour}c, red line). After
that, for $n>n_{min}$, the free energy increases again because
$\alpha$-helical oligomers are not stable under these conditions. The
existence of a local minimum of this type has never been observed in
atomic or molecular systems where the nucleation barrier increases
monotonically until a critical size, and then decreases monotonically
for larger sizes \cite{kelton91}.  Inclusion of all $m$ values sampled
during simulation in the summation relieves all constrains on the
number of $\beta$-sheets formed within the oligomer and we have found
that the position of the local minimum of $F_2$ moves to a larger
value $n_{min}\sim 15$ (Fig.~\ref{fig:contour}c, green line).

These results provide a molecular description of the origin of the
coupling between the nucleation events leading to the formation of
fibrillar structures. The ability of polypeptide chains to reorder
within oligomers into a fibrillar structure stabilises oligomeric
assemblies since the surface of a growing $\beta$-sheet acts as a
substrate for the attachment of other $\alpha$ helical peptides. This
self-templated nucleation mechanism also exists for the aggregation
process above the folding temperature (see Fig.~\ref{fig:snapshot}),
but the associated nucleation barriers are smaller. In systems where
proteins form complex structural motifs in the monomeric phase, the
activation barriers for $\beta$-sheet aggregation are likely to be
higher, and the templating effect in the nucleation mechanism might be
more pronounced. 
A better understanding of this mechanism should lead to an increasing ability
to modulate the growth of peptide and protein aggregates, and it
should play an important role in the development of therapies for
conditions such as Alzheimer's and Parkinson's
diseases \cite{haass07}.

We thank A. Aggeli, E. Paci, P. Olmsted, and J. R. Banavar for
illuminating discussions. Corresponding authors: s.auer@leeds.ac.uk
and mv245@cam.ac.uk


\end{document}